\RequirePackage{fix-cm}

\documentclass[smallextended]{svjour3}       
\smartqed{} 
\usepackage{graphicx}
\usepackage[hyphens]{url}
\usepackage[normalem]{ulem} 
\usepackage{tabularx} 
\usepackage{booktabs} 
\usepackage[toc, page]{appendix} 

\usepackage[dvipsnames]{xcolor}
\usepackage{soul}

\usepackage{tablefootnote}

\usepackage{tcolorbox} 
\usepackage{textcomp} 

\renewcommand{\textrightarrow}{$\rightarrow$}
\definecolor{hzcolor}{RGB}{10, 186, 181}

\usepackage{soul}
\usepackage{float} 
\usepackage{enumitem} 

\tcbuselibrary{many}
\newtcolorbox{rqsumbox}[1][]{%
    enhanced,
    left=4pt,
    right=4pt,
    top=4pt,
    bottom=4pt,
    colback=gray!5,
    colframe=gray!40!black,
    attach boxed title to top left={
        xshift=0.5cm,
        yshift=-\tcboxedtitleheight/2},
    top=4mm,
    coltitle=black,
    before skip=6pt,
    after skip=6pt,
    #1
}

\usepackage{graphics}
\usepackage{subcaption} 
\usepackage{enumitem}
\usepackage{todonotes}
\if@todonotes@disabled

\else

\fi
\usepackage{cite}
\definecolor{wzcolor}{HTML}{8D608C}%

\newcommand{\rheader}[1]{\bigskip \noindent\textbf{#1} }

\makeatletter
\usepackage{csquotes}
\usepackage{hyperref}
\newcounter{expcounter} 
\newcommand{\refexample}[3]{
    \noindent\stepcounter{expcounter}\phantomsection{}%
    \def\@currentlabel{\theexpcounter{}}%
    \begin{tcolorbox}[sharp corners, title=Example \theexpcounter{}\label{#1}]
    #2 \\

    Comment:
    \begin{displayquote}
        \textit{#3}
    \end{displayquote}%
    \end{tcolorbox}
}
\makeatother

\newcommand{\etal}{et~al.} 

\definecolor{ahcolor}{HTML}{FF7500}
\definecolor{ahhlcolor}{HTML}{FAFF72}

\newcommand{\Qwc}{Q\textsubscript{\textit{chat}}} 
\newcommand{\Qd}{Q\textsubscript{\textit{disc}}} 
\newcommand{\Qnd}{Q\textsubscript{\textit{nd}}} 
\newcommand{\Qa}{Q\textsubscript{\textit{a}}} 
\newcommand{\Qdaa}{Q\textsubscript{\textit{d/aa}}} 
\newcommand{\Qda}{Q\textsubscript{\textit{d/a}}} 
\newcommand{\Qhda}{Q\textsubscript{\textit{hd/a}}} 
\newcommand{\tfa}{t\textsubscript{FA}} 
\newcommand{\taa}{t\textsubscript{AA}} 

\begin{document}

\title{An empirical study of question discussions on Stack Overflow}

\author{Wenhan Zhu \and
        Haoxiang Zhang  \and \\
        Ahmed E. Hassan \and
        Michael W. Godfrey
}

\institute{Wenhan Zhu $\cdot$ Michael W. Godfrey
    \at{Software Analytics Group (SWAG) \\
    University of Waterloo\\
    Waterloo, ON, Canada
} \\
    \email{\{w65zhu, migod\}@uwaterloo.ca} \\
    \and
    Haoxiang Zhang
    \at{ Centre for Software Excellence at Huawei, Canada
} \\
    \email{haoxiang.zhang@acm.org} \\
    \and
    Ahmed E. Hassan
    \at{ Software Analysis and Intelligence Lab (SAIL)\\
    	Queen's University\\
    	Kingston, ON, Canada
    } \\
    \email{ahmed@cs.queensu.ca} \\
}

\date{Received: date / Accepted: date}

\maketitle

\begin{abstract}

Stack Overflow provides a means for developers to exchange knowledge. While much previous research on Stack Overflow has focused on questions and answers (Q\&A), recent work has shown that \emph{discussions} in comments also contain rich information. On Stack Overflow, discussions through comments and chat rooms can be tied to questions or answers. In this paper, we conduct an empirical study that focuses on the nature of question discussions. We observe that: (1) Question discussions occur at all phases of the Q\&A process, with most beginning before the first answer is received. (2) Both askers and answerers actively participate in question discussions; the likelihood of their participation increases as the number of comments increases. (3) There is a strong correlation between the number of question comments and the question answering time (i.e., more discussed questions receive answers more slowly). Our findings suggest that question discussions contain a rich trove of data that is integral to the Q\&A processes on Stack Overflow. We further suggest how future research can leverage the information in question discussions, along with the commonly studied Q\&A information.

\end{abstract}





\maketitle

\section{Introduction}


Stack Overflow is a technical question answering (Q\&A) website widely used by developers to exchange programming-related knowledge through asking, discussing, and answering questions.
The Q\&A process on Stack Overflow creates a crowdsourced knowledge base that provides a means for developers across the globe to collectively build and improve their knowledge on programming and its related technologies.
Stack Overflow has become one of the largest public knowledge bases for developers with more than 21.9 million questions as of December 2021~\cite{stackoverflow2021datadump}.
A survey shows that retrieving information from Stack Overflow is an essential daily activity for many software developers~\cite{7498605}. 


On Stack Overflow, users can ask, answer, and discuss questions, and each question can receive multiple proposed answers. 
The user who asked the question (i.e., the ``asker'') can decide to mark one answer as \emph{accepted}, indicating that it resolves their question authoritatively.
While ultimately Q\&A is the most important activity on Stack Overflow, users can also post comments and/or start chat rooms that are tied to a specific \emph{post} (i.e., question or answer). 
In this paper, we refer to comments and chat rooms messages on Stack Overflow as \textit{discussions}; each discussion is associated with a single question (a \emph{question discussion}) or proposed answer (an \emph{answer discussion}).


Researchers have extensively studied the questions and answers on Stack Overflow. 
These studies ranged from finding out common question types~\cite{allamanis2013and} to predicting the best answers~\cite{tian2013towards}. 
The Q\&A processes on Stack Overflow are commonly viewed as two independent events by the studies. 
The first event is asking the question; this occurs when a user posts a question on the platform. 
The second event is answering the question; this normally occurs when another user posts an answer to a question. 
However, commenting as a communication channel allows for user interactions beyond simple asking and answering.
A recent study has shown that comments can be helpful in the Q\&A process by providing support knowledge, such as code examples, references, and alternative suggestions~\cite{sengupta2020learning}, whereas previous research has focused primarily on answer comments.
Some studies leverage answer comments to study the quality of answers on Stack Overflow.
For example, Zhang~\etal{}~\cite{8669958} leveraged comments highlighting obsolete information regarding their associated answers.
As a Q\&A platform, most content on the platform is consumed by viewers long after the question is answered.
If misleading information exists on the platform, it can convey false information within the community.
Another study~\cite{chen2019reliable} used comments as a sign of whether the community is aware of the security vulnerabilities contained in the answer.
Meanwhile, some studies have also focused on the presentation of knowledge on Stack Overflow.
These studies also approach the issue from the answer perspective.
One study~\cite{zhang2019reading} highlights that while users are reading answers on Stack Overflow, they should not ignore the information contained in their associated comments.
In their next study\cite{zhang2021comments}, they showed that the current mechanisms on Stack Overflow to display comments is not ideal and can hurt the users when they are reading answers.

In our study, we focus on question comments.
More specifically, we theorize that the commenting activities forms a discussion and our focus is to understand how the discussions affects the Q\&A process on Stack Overflow.
Unlike previous studies that mostly focus on answer comments which occur after a question has been answered, our study focuses on question comments which can occur before the question is answered.

\begin{figure}[ht]
    \centering
    \includegraphics[width=0.9\textwidth]{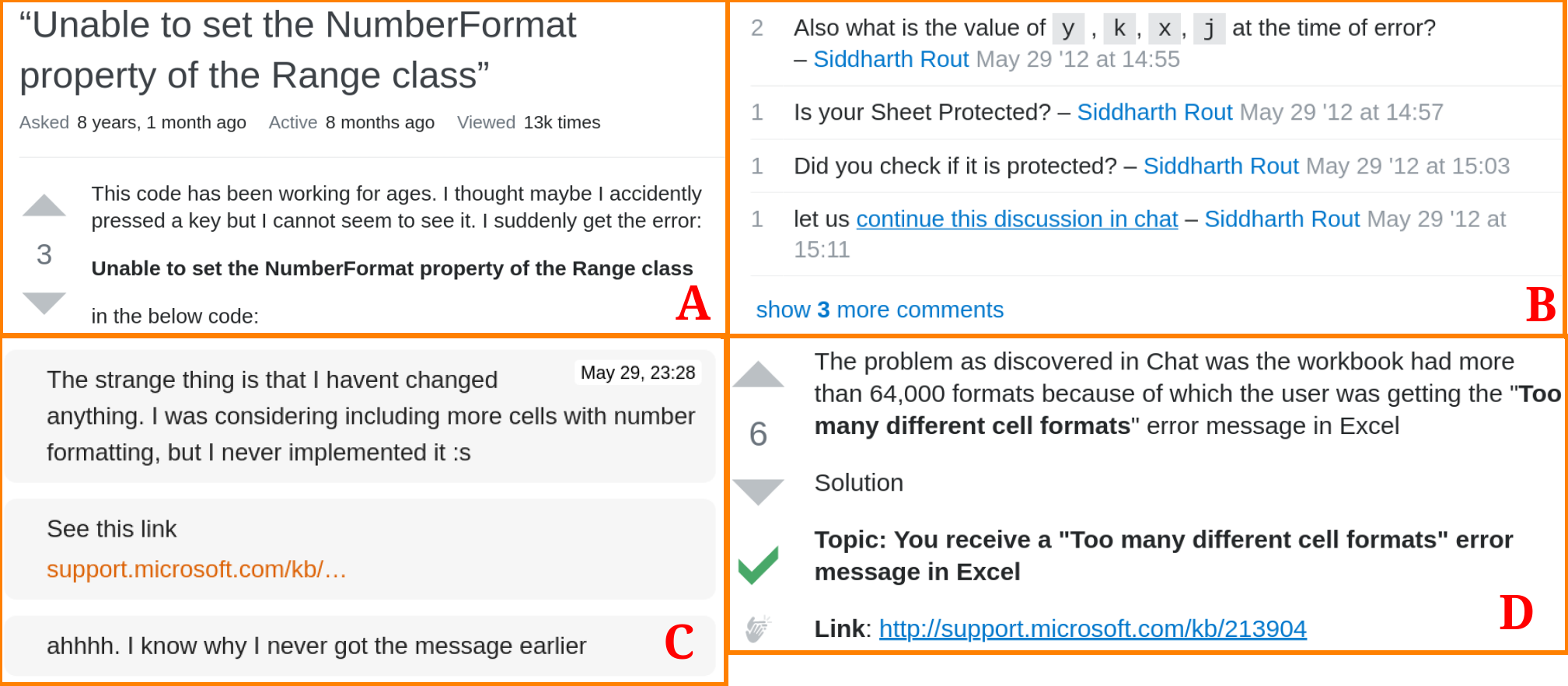}
    \caption{An example of the Q\&A process involving discussions: (A) a
	user (the ``asker'') asked a question; (B) another user (the
	``answerer'') started discussing with the asker in the comment
	thread; (C) the question was further clarified then resolved in the
	chat room; (D) the content of the comments and chat messages that
	led to the resolution of the question were summarized as an answer,
	which was marked as the accepted answer by the asker.
    }\label{fig:example_question}
\end{figure}

To help understand why it is important to study how question discussions integrate with the Q\&A process, we now consider a motivating example.
Fig.~\ref{fig:example_question} shows a question titled \emph{``Unable to set the \texttt{NumberFormat} property of the \texttt{Range} class\footnote{\url{https://stackoverflow.com/questions/10801537/}}.''}
Four minutes after the question was asked, another user posted a comment --- attached to the question --- asking for clarification on the problematic code snippet. 
A chat room was then created for the asker and the user to continue the discussion in real-time.   
A consensus was reached in the chat, and the results were summarized and posted as a proposed answer by the user, which the asker designated as \emph{accepted}.
This example highlights how the process of asking and answering questions is enabled by the discussion mechanisms of commenting and chatting, allowing a resolution to be reached quickly.
That is, the question discussion can serve as a simple and effective socio-technical means to achieving closure on the question.

In this work, we use the Stack Overflow data dump from December
2021~\cite{stackoverflow2021datadump} as our dataset; this dataset contains
43.6 million comments and 1.5 million chat messages.  We use this data to
explore the nature of question discussions and how they integrate with the
crowdsourced Q\&A process on Stack Overflow.  To make our study easy to follow,
we use the following notations to refer to different groups of questions
observed within the dataset:

\vspace*{.5em}
\begin{tabularx}{.95\textwidth}{lXr}
\toprule
\emph{Symbol} & \emph{Meaning}			& \emph{\# in dataset} \\
\midrule
\Qd{}         & Questions with comments		& 13.0 M	    \\
\Qwc{}        & Questions with chat rooms (and comments)  & 27,146 \\
\Qnd{}        & Questions with no discussions	& 9.0 M	    \\
\Qa{}         & Questions with answers		& 18.8 M    \\
\Qda{}        & Questions with both discussions and answers	& 10.5 M	\\
\Qdaa{}       & Questions with both discussions and accepted answers & 6.1 M \\
\Qhda{}       & Questions with both discussions with ``hidden
		comments''\footnotemark{} and answers		& 1.6 M \\
\bottomrule
\end{tabularx}

\footnotetext{In Stack Overflow, comments are ``hidden'' (i.e., elided from
view) by default when there are six or more attached to the same question.}

\vspace*{1em}

\noindent
Specifically, we investigate and answer three research questions (RQs):
\begin{description}
    \item{\textbf{\nameref{text:rq1}}} \\[1em]
    We found that question discussions occur in 59.2\% of the questions on Stack Overflow. 
    More specifically, 13.0 million questions have comments (i.e., \Qd{}) with a median of 3 comments, and 27,146 questions have chat rooms (i.e., \Qwc{}).
    The popularity of question discussions is also increasing, with the proportion of questions with discussions nearly doubling from 32.3\% in 2008 to 59.3\% in 2018.
    Question discussions exist in all phases of the Q\&A process on Stack Overflow. 
    In questions that are both discussed and have an accepted answer (i.e., \Qdaa{}), discussions in 80.6\% of the questions begin before the accepted answer was posted.
    We found that the duration of question discussions can extend beyond the Q\&A process: In 28.5\% of \Qdaa{}, question discussions begin before the first answer and continue after the accepted answer is posted; and in 19.4\% of \Qdaa{}, question discussions begin after the question receives its accepted answer.
\vspace*{1em}

\item{\textbf{\nameref{text:rq2}}} \\[1em]
    We found that 16.0\% (i.e., 2.6 million) of registered users on Stack Overflow have participated in question discussions, which is comparable to the number of users who have answered questions (i.e., 16.7\%).
    Question discussions allow askers and answerers to communicate with each other directly, enabling fast exchanges on the issues of concern.
    For questions that have both discussions and answers (i.e., \Qda{}), we found that as the number of comments increases, both askers and answerers were more likely to participate in the question discussions.
    Also, we found that when there are six or more comments present (i.e., \Qhda{}), then there is a high likelihood of both askers (90.3\%) and answerers (51.9\%) participating in the discussions.
\vspace*{1em}

\item{\textbf{\nameref{text:rq3}}} \\[1em]
    Question discussions tend to lead to more substantial updates to the body of the original question. 
    For example, a median of 114 characters are added to the question body when the question discussion has a chat room instance (i.e., \Qwc{}). 
    While most other questions have no change in their question body length, a larger proportion of questions with comments are revised, with an increase in the question body length compared to questions with no discussion.
    Questions with more comments receive answers more slowly, with a Spearman correlation of $\rho = 0.709$ between the number of comments and the \emph{answer-receiving-time} for the first answer. 

\end{description}

The main contribution of our study is to highlight that discussions are an integral part of the Q\&A process on Stack Overflow.
Compared to the common assumptions that asking and answering questions are separate events in many studies, our work suggests that a large proportion of questions on Stack Overflow are answered after interactions between askers and answerers in question discussions.
Our study suggests that question discussions is a very common activity comparable to answering activity on Stack Overflow.
Question discussions have a high active user base (i.e., 16.0\% of active users), and are also comparable to answering (i.e., 16.7\% of active users).
We also observed a strong correlation between the number of comments and the question answering speed, suggesting that question discussions have an impact on creating answers.
Our findings suggest that question discussions can facilitate the Q\&A process since they provide a means for askers and potential answerers to communicate throughout the Q\&A process.
We encourage future research on Stack Overflow to consider question discussions in addition to leveraging the information in the questions and answers of Stack Overflow.

\textbf{Paper Organization.} 
The rest of this paper is organized as follows. 
Section~\ref{sec:background} introduces Q\&A on Stack Overflow and commenting/chatting on Stack Overflow.
Section~\ref{sec:data_collection} describes how we collect data for our analysis. 
Section~\ref{sec:results} details the results of our empirical study. 
Section~\ref{sec:implications_and_discussions} discusses our findings and their implications.
Section~\ref{sec:threats_to_validity} describes threats to the validity of our study. 
Section~\ref{sec:related_work} surveys related research. 
Finally, Section~\ref{sec:conclusion} summarizes the findings of our study.


\section{Background}\label{sec:background}

\subsection{The Q\&A Process on Stack Overflow}

Stack Overflow is a technical Q\&A website where users ask, answer, and discuss questions related to programming and software development.
Stack Overflow has been widely embraced by the software engineering community, and has become the largest public knowledge base for programming-related questions.  
There are 21.9 million questions together with 32.7 million answers on Stack Overflow as of December 2021.

The Stack Overflow Q\&A process begins with a user posting a \emph{question} that relates to programming or a similar technical topic.  
At that point, other users can start to engage either by proposing an \emph{answer}, or by taking part in a \emph{discussion} in the form of a \emph{comment} or a \emph{chat room}.
Discussions can be attached to either the original question (i.e., a \emph{question discussion}) or one of the proposed answers (i.e., an \emph{answer discussion}).  
If a proposed answer successfully resolves the question, the user who asked the original question (i.e., the \emph{asker}) may at their discretion choose to designate that answer as the \emph{accepted answer}.  
Once an accepted answer has been selected, users may continue to contribute to the question thread by adding new answers or editing existing content; in practice, however, user activity related to that question and its answers tends to diminish sharply at that point~\cite{DBLP:conf/msr/BaltesDT008}.
We note that the Stack Overflow uses the term \emph{post} internally to refer to either a question or answer, but not a discussion.

\subsection{Discussions on Stack Overflow}

In this work, we focus on question discussions to better understand how discussions affect the crowdsourced knowledge sharing activities once a question is posted, especially those that occur early in the Q\&A process.

Stack Overflow offers two different forms of communication channels for users to discuss on questions and answers, that is, commenting as an asynchronous communication channel and chatting as a synchronous communication channel.
When users are commenting, they may not expect an immediate reply.
Meanwhile, when users are chatting, a live session is formed where information flows freely within the group in real-time~\cite{7498605}.
On Stack Overflow, users begin discussions in comments.
When extended discussions occur in comments, users are proposed with continuing the discussions in dedicated chat rooms.
While commenting is the dominating communication channel on the Stack Overflow for question discussions, whenever possible, we take special notice of the existence of chat rooms since they represent a different form of communication channel compared to comments.

As previously mentioned, users can attach comments to a post (i.e., a question or answer).
Stack Overflow considers comments as \textit{``temporary `Post-It' notes
left on a question or
answer.''}\footnote{\label{ft:so_comment}\url{https://stackoverflow.com/help/privileges/comment}}
Stack Overflow users are encouraged to post comments \textit{``to request clarification from the author; leave constructive criticism to guide the author in improving the post, and add relevant but minor or transient information to a post.''} 
When multiple comments are present in the same post, they form a \emph{comment thread}.


Stack Overflow offers \textit{real-time, persistent collaborative chat for the community}\footnote{\url{https://chat.stackoverflow.com/faq}} with chat rooms. 
Stack Overflow promotes users to continue the discussions in a chat room when there are more than three back-and-forth comments between two users (i.e., at least 6 in total). 
Users are prompted with a message before a chat room can be created: \textit{``Please avoid extended discussions in comments.  Would you like to automatically move this discussion to chat?''} When the user agrees to create the chat room, an automated comment is posted and contains a link to the newly created chat room.
In the newly created chat room, automated messages are posted indicating the associated question and the comments leading to the chat room. 
Users can also create chat rooms directly that are not associated with questions or answers.


\section{Data Collection}\label{sec:data_collection}

In our study, we use the Stack Overflow data dump from December 2021~\cite{stackoverflow2021datadump}.
The data dump is a snapshot of the underlying database used by Stack Overflow; it contains all meta-data for each comment, such as which user posted the comment and which question the comment is associated with. 
We mainly used the \texttt{Posts} and \texttt{Comments} table from the dataset to extract the required information.
The data dump also contains the history of each question, via the \texttt{PostHistory} table.
We analyze the history of each question to reconstruct the timeline of when the question was created, edited, commented, and answered.

\begin{figure}[ht]
    \centering
    \includegraphics[width=\textwidth]{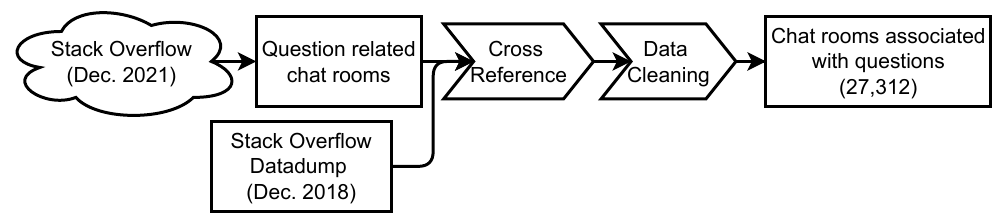}
    \caption{An overview for the creation of \Qwc{} (questions with chat rooms)}\label{fig:questionCollection}
\end{figure}

Data about chat rooms is not contained in the Stack Overflow data dump; instead, we collected it manually by crawling the Stack Overflow website itself\footnote{We've made our dataset open access on Zenodo: \url{https://zenodo.org/record/5516190}}.  
We also labelled the chat room instances based on whether they are \emph{general}\footnote{General chat rooms are standard chat rooms on Stack Overflow that are not associated with a question or an answer.}, attached to a \emph{question}, or attached to an \emph{answer}.
After cross-referencing their associated question IDs with the Stack Overflow data dump, we removed chat room discussions that are unrelated to programming, such as those on Meta Stack Overflow which focuses on the operation of Stack Overflow itself.  
This left us with a total of 27,312 chat rooms comprising 1.5 million messages that are associated with 27,146 questions as of December 2021. 
Figure~\ref{fig:questionCollection} shows the detailed extraction process of chat rooms from Stack Overflow. 

\section{Case Study Results}\label{sec:results}

In this section, we explore the underlying motivation, the approach taken,
and the results of our three research questions (RQs) concerning question discussions on Stack Overflow.  

\subsection{\textbf{RQ1:} 
    How prevalent are question discussions on Stack Overflow?}
\label{text:rq1}

\rheader{Motivation:} 
As a technical Q\&A platform related to programming, Stack Overflow hosts a large number of questions~\cite{Treude:2011:PAA:1985793.1985907}.  
From the user's point of view, creating an answer can be challenging since the initial version of a question is often incomplete or ambiguous.  
For this reason, potential answerers may first wish to engage the asker in a discussion to clarify their intent and possibly seek additional context, which is typically done using comments attached to the question.  
If the discussion proves to be fruitful, the user may then post an answer based on the discussion; also, the asker may decide to edit the original question to clarify the intent for other readers.
For example, Example~\ref{exp:point_out_problem} shows a comment pointing out an confounding issue in the original question.  
After the discussions, the asker acknowledged the issue and edited the original question for clarity.

A prior study showed that active tutoring through discussions in chat rooms can substantially improve the quality of newly posted questions by novice users~\cite{ford2018we}.  
However, it is labor intensive to provide such tutoring with an average of more than 7,000 new questions posted per day on Stack Overflow in 2019. 
At the same time, there has been no detailed study of question discussions as yet; in this RQ, we explicitly study question discussions to gain a better understanding of their prevalence in the Q\&A process.

\refexample{exp:point_out_problem}{In a comment linked to a question
titled: ``Write to Excel --- Reading CSV with Pandas \& Openpyxl -
Python.\footnotemark'', a user observed that the example CSV file given in
the question did not follow the CSV standard, and suggested the asker to
double check the input format.}{The structure of the first three lines
doesn't match the structure of lines 5 onwards so you cannot read this file
with a CSV library. Please check the provenance of the file and what it
should look like. I suspect you probably want to skip the first four
lines.}

\footnotetext{\url{https://stackoverflow.com/questions/48956597/}}

\rheader{Approach:}%
We begin our study of the prevalence of question discussions by investigating the trend in the number and proportion of question discussions over the years. 
We distinguish between answered questions with and without an accepted answer to investigate whether there exists a difference between the two groups of questions.
\begin{figure}[ht] \centering
\includegraphics[width=\textwidth]{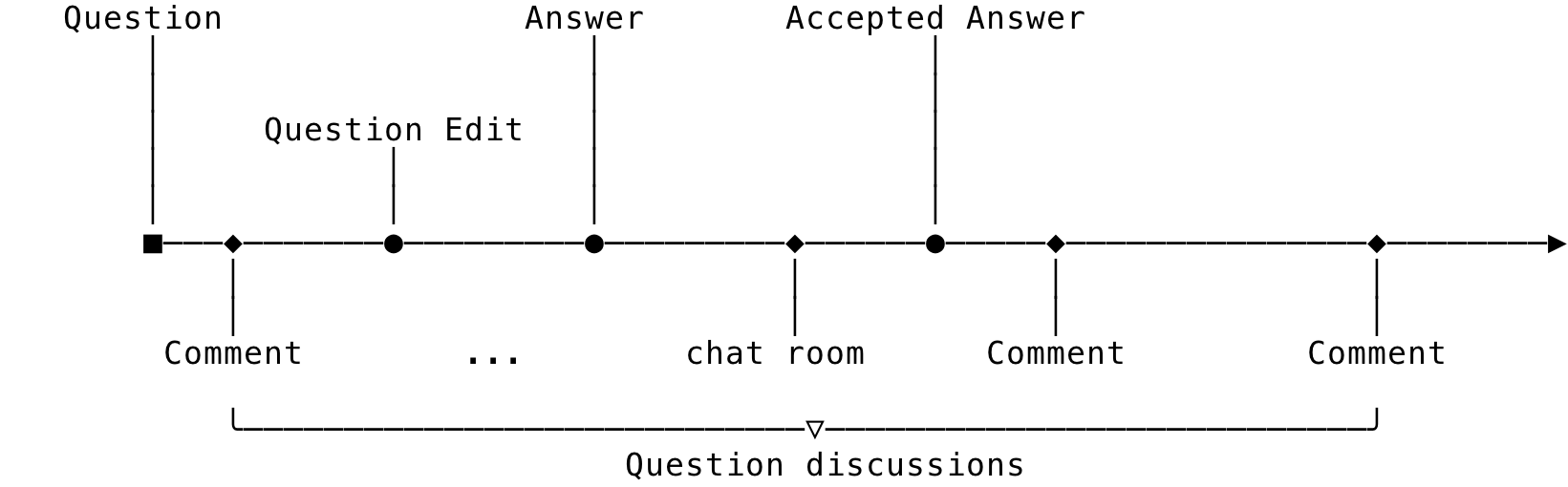}
\caption{Timeline of question thread events. Question discussions can occur at any time since the creation of a question.}\label{fig:question_timeline}
\end{figure}

We then study when question discussions occur relative to key events in the Q\&A process. 
After a question is posted on Stack Overflow, several different types of follow-up events may occur, as illustrated by Fig.~\ref{fig:question_timeline}. 
For example, after a question is posted any of the following can occur:
\begin{itemize}
\item other users can propose answers to the question;
\item users can post comments to discuss either the question or the
    associated answers;
\item the asker can mark one of the answers as \emph{accepted}; and
\item the question (and proposed answers) can be edited for clarity.
\end{itemize}

For each question, we construct the timeline consisting of each event, and we analyze the prevalence of question discussions with respect to other Q\&A activities. 
Here, we focus mainly on two key events: when the question receives its first answer, and when it receives the accepted answer. 

\begin{figure}[ht] \centering
\begin{subfigure}[b]{\textwidth}
    \includegraphics[width=\textwidth]{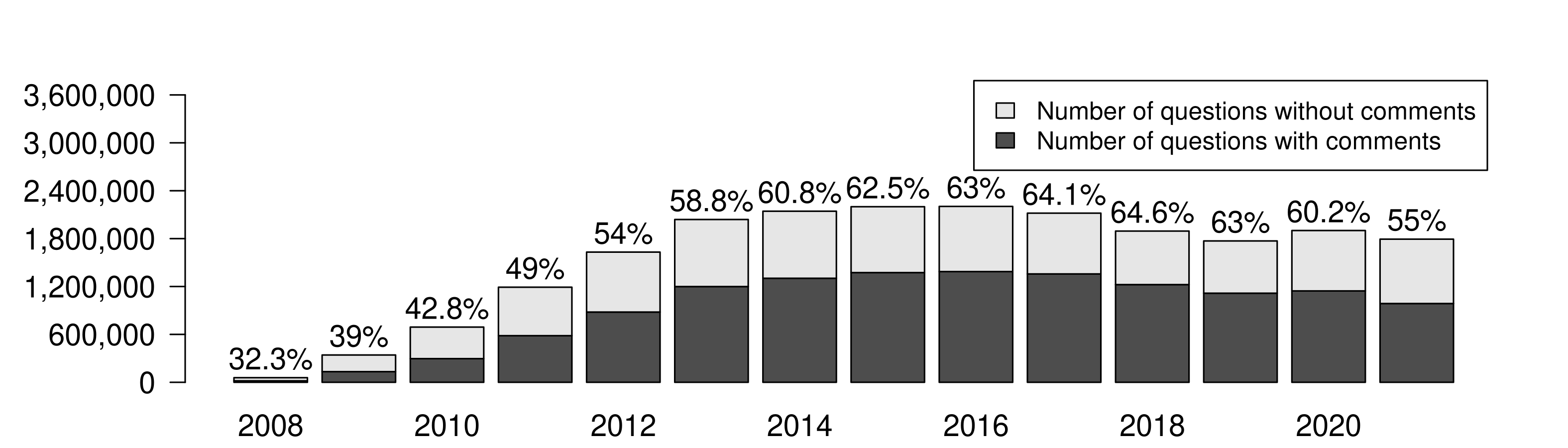}
    \caption{All questions}\label{fig:question_discussion_year_dist_all} 
\end{subfigure}

\begin{subfigure}[b]{\textwidth}
    \includegraphics[width=\textwidth]{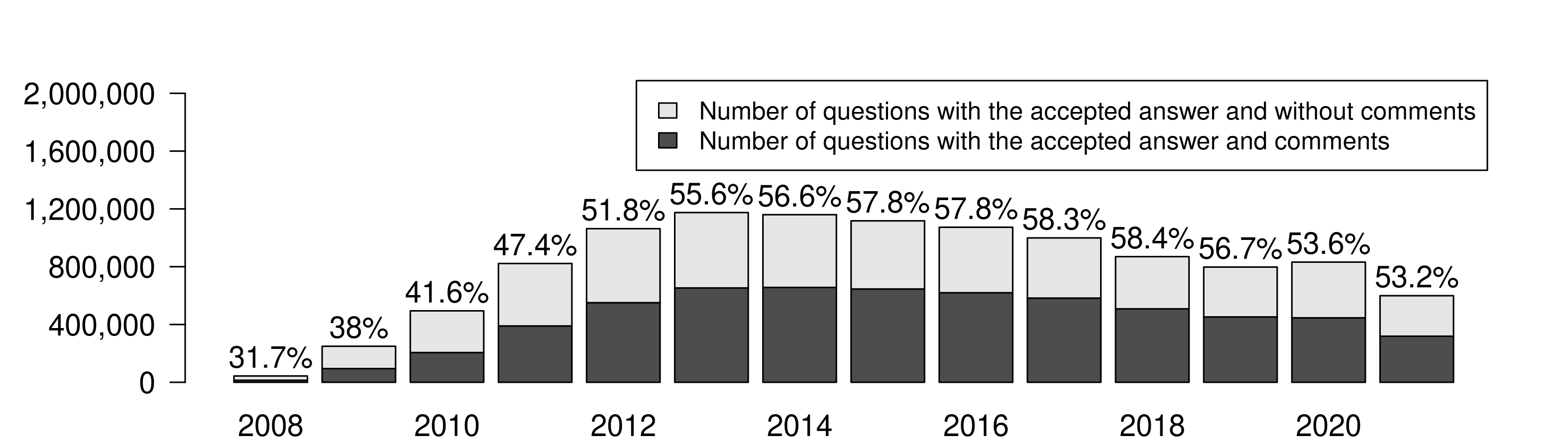}
    \caption{Questions with the accepted answer}\label{fig:question_discussion_year_dist_accepted} 
\end{subfigure}

\caption{The number and proportion of questions with comments}
\end{figure}

\rheader{Results:}%
\textbf{Stack Overflow questions are discussed by 43.6 million comments and 1.5 million chat messages, forming a large dataset of community question discussions, in addition to the 22.0 million questions and 32.7 million answers.} 
The proportion of questions with discussions also nearly doubled from 32.3\% in 2008 to 59.3\% in 2013, and has remained roughly stable since then. 
Fig.~\ref{fig:question_discussion_year_dist_all} shows the number and
proportion of questions with discussions per year, and
Fig.~\ref{fig:question_discussion_year_dist_accepted} suggests a similar trend for questions with an accepted answer. 
Since a question may receive its first comment several years later, it is likely that the proportion of recent years will increase slightly in the future.


\begin{figure}[ht] \centering
    \includegraphics[width=\textwidth]{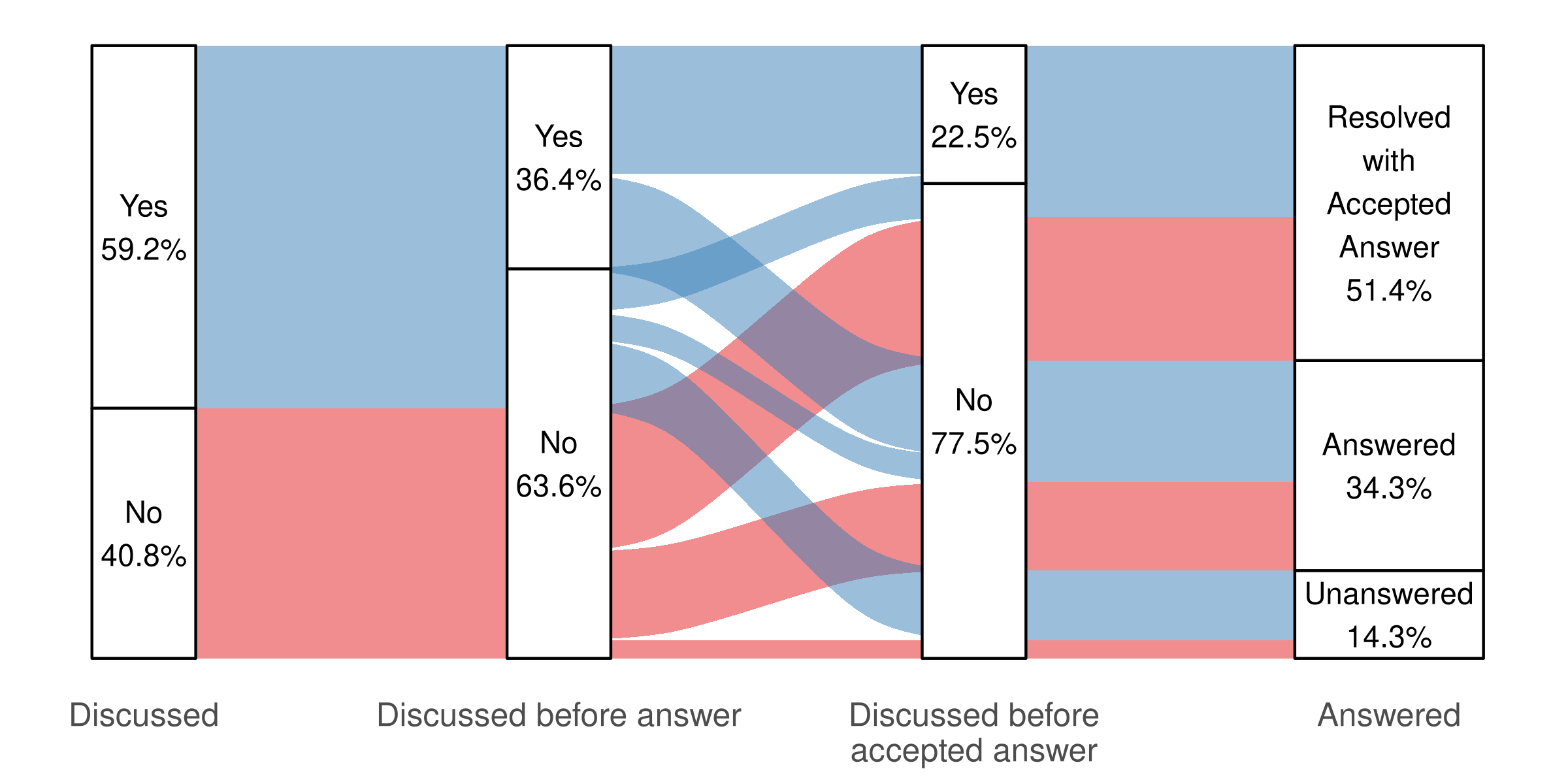}
\caption{Question discussion with respect to answering events during the Q\&A process. The blue bands represent questions with discussions and the red bands represent questions without discussions.}\label{fig:question_answer_discussion_flow} \end{figure}

\textbf{Question discussions occur throughout the Q\&A process, ranging from before the first answering event to after the accepted answer is posted.}
Fig.~\ref{fig:question_answer_discussion_flow} shows the proportion of question discussions relative to answering events in the Q\&A process.
The height of the band across each vertical line indicates the proportion of questions with a specific activity occurring in that phases of a question thread's life cycle.
For example, from the left-most bar, all questions can be split into two groups: questions with discussions (\Qd{}) and questions without discussions (\Qnd{}).
The top band (with strata in blue) represents 59.2\% of the questions with
discussions and the bottom band (with strata in red) represents 40.8\% of the questions without any discussions.
Flowing from left to right, the strata in blue and red continue to represent the questions with and without discussions until the right most band where it represent the final answering status of the question.

\textbf{In \Qda{}, 76.2\% (i.e., 8.0 million) of the question discussions begin before the first answer is posted, suggesting an influence of question discussions on answering activities.} 
Furthermore, 80.6\% (i.e., 4.9 million) of the question discussions begin before the accepted answer is posted, indicating a slightly more active involvement of question discussions in \Qdaa. 
In answered and solved questions of \Qwc{}, 59.1\% (i.e., 12,507) of the chat activities begin before the first answer is received, and 72.9\% (i.e., 10,172) of the chat activities begin before the accepted answer is posted. 

The early occurrence of question discussions in the Q\&A process suggests that they enable interested users to engage with the asker informally, to allow for clarification.
For example, in Ex.~\ref{exp:qa_start_from_discussion}, 13 minutes after the question was initially posted, a user asked for a concrete example that can demonstrate the problem the asker had. 
The asker then updated the question with the requested information. 
The question was answered 15 minutes later, incorporating the newly added information based on the discussions.

\refexample{exp:qa_start_from_discussion}{A user comments to ask for information in a question titled ``Can I modify the text within a beautiful soup tag without converting it into a string?\footnotemark''}{UserB: Please give an example html that demonstrates the problem. Thanks. \hfill \textnormal{[2014-09-16 13:15]} \\
UserA (the asker): Just added some example html, sorry about that. \hfill \textnormal{[2014-09-16 13:20]}} %

\footnotetext{\url{https://stackoverflow.com/questions/25869533/}}

\textbf{In 28.5\% (i.e., 1.7 million) of \Qdaa{}, the discussions begin before the accepted answer has been received, and continue after the accepted answer is posted. 
Furthermore, 19.4\% (i.e., 1.2 million) of the question discussions begin after the accepted answer is posted.} 
These findings indicate that the community may continue to discuss questions even after the asker has designated a ``best'' answer that solves their problem~\cite{anderson2012discovering}. 
This may be due to the fact that software development technologies tend to evolve rapidly; old ``truths'' may need to be updated over time, and additional discussions may provide new insights despite the asker considering the question to be solved.
Example~\ref{exp:comment_warning} shows a comment that pointed out a potential security vulnerability in the code snippet 5 years after the initial question is posted.

\refexample{exp:comment_warning}{A user posted a comment to warn about a
potential security vulnerability 5 years after a question was
posted.\footnotemark''}{Beware.  If you've configured your Struts
application in this particular way (setting `alwaysSelectFullNamespace' to
`true'), your application is very likely vulnerable to CVE-2018-11776:
\url{semmle.com/news/apache-struts-CVE-2018-11776} }

\footnotetext{\url{https://stackoverflow.com/questions/17690956/}}

\begin{rqsumbox}[title={RQ1 Summary:}]
There are 44.6 million comments and 1.5 million chat room messages in our dataset, which forms a large corpus of question discussion activities on Stack Overflow. 
Since the introduction of comments, the popularity of question discussions has nearly doubled from 32.3\% in 2008 to 59.3\% in 2013 and has remained stable since.
The occurrence of question discussions is prevalent throughout the Q\&A process.
While question discussions in most questions (76.2\% in \Qda{} and 80.6\% in \Qdaa{}) begin before the answering activities, question discussions can continue or even begin after the accepted answer is posted.
\end{rqsumbox}


\subsection{\textbf{RQ2:} How do users participate in question discussions?}\label{text:rq2}

\rheader{Motivation:}%
The crowdsourced Q\&A process on Stack Overflow is driven by user
participation. In addition to the questions and answers, question discussions
are also part of the user-contributed content on Stack Overflow. In this RQ, we
explore how different users participate in question discussions, to better
understand how question discussions facilitate the Q\&A process. 

We focus on two aspects of user participation. First, we investigate the
overall user participation in question discussions on Stack Overflow. We note
that in RQ1, we observed a high proportion of questions with discussions; here,
we focus on the users who participate in question discussions. Second, we
change the scope to focus on the question-level discussion participation. We are
interested in what other activities that the participating users join in on. For
example, did the user ask the question in the first place, or did the user post
an answer for the question.

\rheader{Approach:}%
To study user participation in question discussions and gain an overall idea of the popularity of discussion activities compared to other activities on Stack Overflow, we extract from the data dump the list of all users who contributed content to Stack Overflow. 
In particular, we sought users who asked, answered, or discussed questions; we note that while other activities, such as voting, may help the community, we do not consider these activities in our study as they do not directly contribute content.
We also ignored activity related to answer discussions, as it was outside of the scope of our investigations.

We extracted the unique \emph{UserIDs} from all questions, answers, and question comments to build the groups of users who participated in each of those
activities.
We then compared the intersection between the different sets of users to determine which of them participated in multiple types of activities on Stack Overflow.

\begin{figure}[ht]
    \centering
    \includegraphics[width=\textwidth]{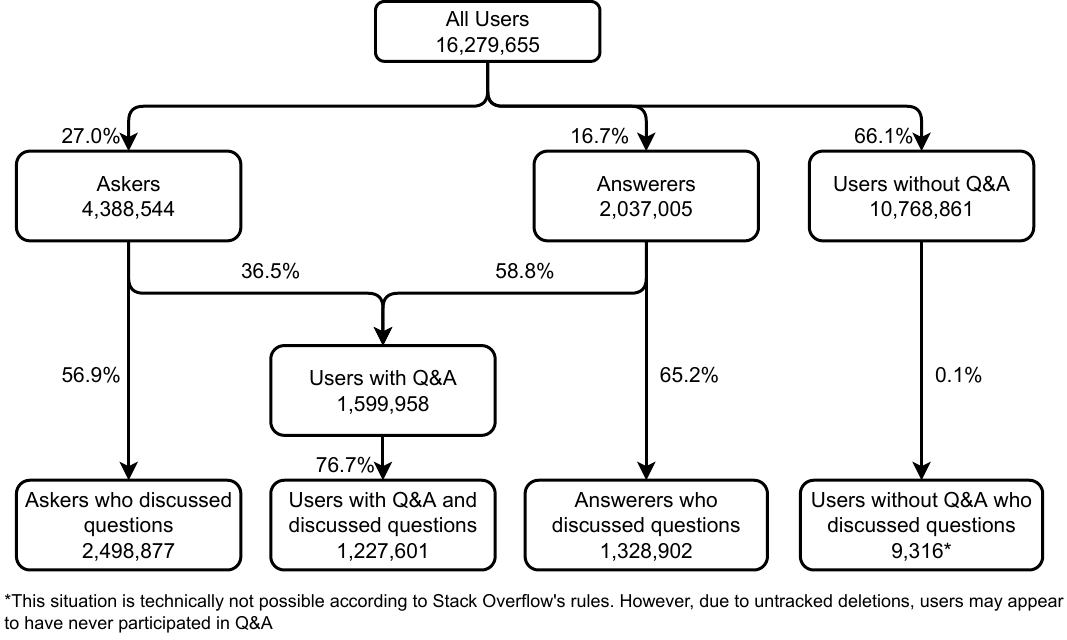}
    \caption{The number of users who participate in different types of activities on Stack Overflow, and the number and proportion of users who participate in question discussions.}\label{fig:active_user_part}
\end{figure}

\rheader{Results:} %
%
\textbf{2.6 million (i.e., 16.0\%) users on Stack Overflow have participated in
question discussions.} Fig.~\ref{fig:active_user_part} shows the overlap of
the number of users participating in different activities on Stack Overflow. We
observe that 95.7\% of users who participated in question discussions also
asked questions on Stack Overflow, and 50.9\% of them answered questions.

%
\textbf{In 60.0\% of \Qda{} (i.e., 7.8 million), askers participate in the question discussions and in 34.1\% of \Qda{} (i.e., 3.6 million), an answerer participated in the question discussion.} The involvement of askers and answerers indicate that the two parties often leverage question discussions as a collaboration medium.

\begin{figure}[ht]
    \centering
    \includegraphics[width=\textwidth]{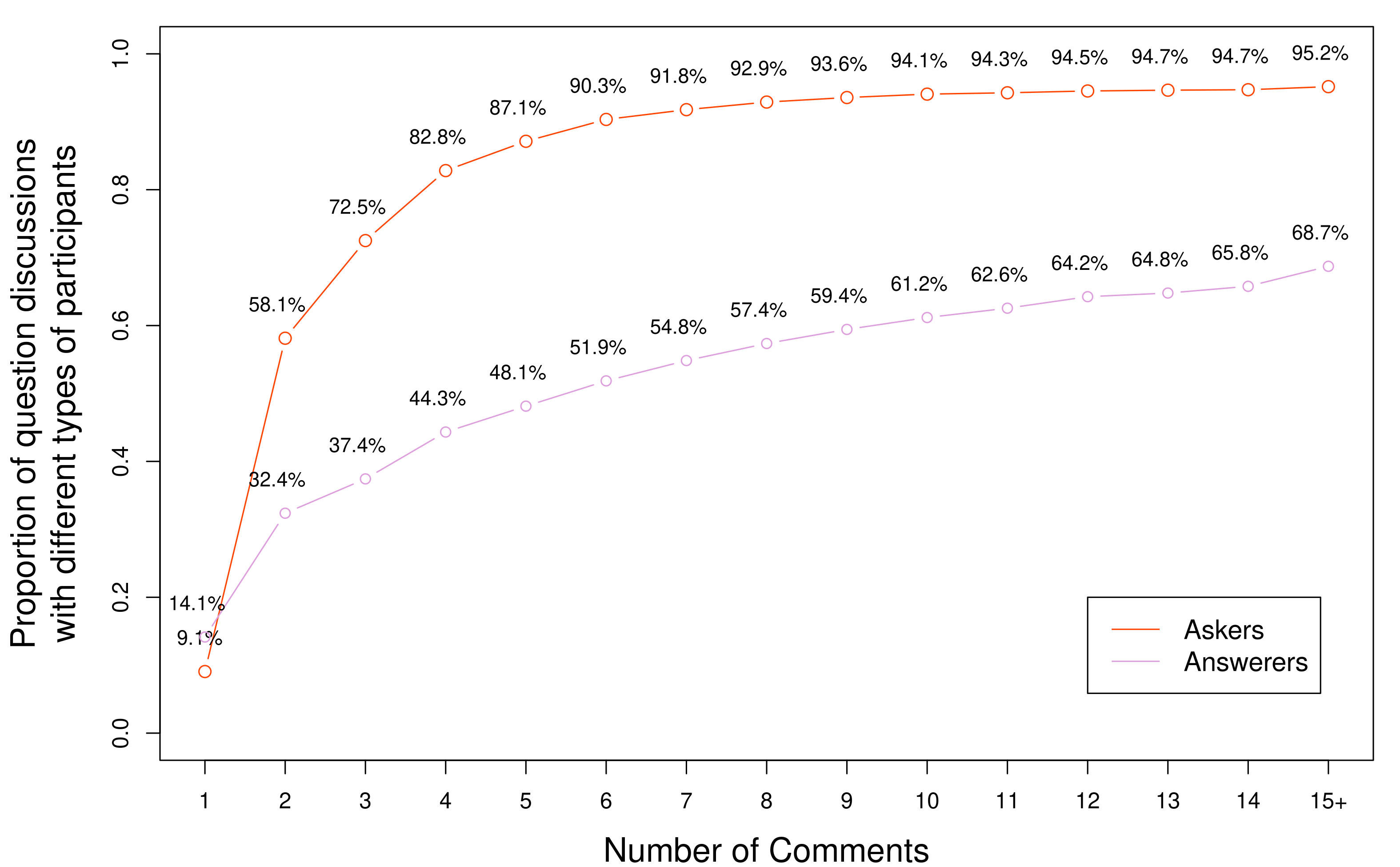}
    \caption{The proportion of question discussions with the participation of askers and answerers}\label{fig:asker_answerer_question_discussion_participation}
\end{figure}

We further investigate the trend of the proportion of questions with askers and answerers in question discussions as the number of comments increases. 
\textbf{When the number of comments increases, a higher proportion of questions have askers and answerers participating}. Fig.~\ref{fig:asker_answerer_question_discussion_participation} shows the trend of the proportion of askers and answerers participating in question discussions as the number of comments increases. When there are at least 6 comments associated with a question (i.e., when Stack Overflow starts to hide additional comments), askers are present in at least 90.3\% of the question discussions and answerers are present in at least 51.9\% of the question discussions.
Moreover, \textbf{when answerers are present in a question discussion, 79.3\% (i.e., 2.8 million) of the answerers and 81.1\% (i.e., 1.5 million) of the accepted answerers joined the question's discussions before posting the answers.}
The increasing proportion and early engagements of answerers in question discussions suggest that users are actively leveraging the question discussions as a communication channel to facilitate the answering of questions.

\begin{rqsumbox}[title={RQ2 Summary:}]
2.6 million (i.e., 16.0\%) users on Stack Overflow have participated in question discussions. These users overlap heavily with users who asked and answered questions on Stack Overflow.
In \Qda{}, 60.0\% of the questions have the asker participating in the question discussion and 34.1\% of the questions have an answerer participating in the question discussion.
The proportion of questions with askers and answerers participating in question discussions increases as the number of comments increases. 
When at least 6 comments are present, more than 90.3\% of the discussions have askers participating and more than 51.9\% have answerers participating.
In 79.3\% of \Qda{} (81.1\% of \Qdaa{}), the answerer (accepted answerer) participated in the question discussion before they posted the answer (accepted answer).
\end{rqsumbox}

\subsection{\textbf{RQ3:} How do question discussions affect the question
answering process on Stack Overflow?}\label{text:rq3}

\rheader{Motivation:} %
On Stack Overflow, questions serve as a starting point for curating crowdsourced knowledge.
Devising a good question can also be a challenging task~\cite{calefato2018ask}.
To encourage users to ask high-quality questions, in late 2019 Stack Overflow modified its reputation system to reward more reputation points on upvotes for questions, increasing the points rewarded from 5 to 10\footnote{\url{https://stackoverflow.blog/2019/11/13/were-rewarding-the-question-askers/}}.
As noted previously, a question can have several follow-up answers; also, discussions can be associated with either the question or its answers.
Questions (and answers) may be edited and revised by their original author, and this happens commonly.\footnote{Comments may be deleted by their author, but they may not be edited in place.}
This may be done to reflect new knowledge learned though the Q\&A process, and to improve the quality of the posts themselves.
In practice, some revisions are editorial or presentational in nature, such as fixing typos and formatting content for readability; however, questions are also edited to improve the quality of the crowdsourced knowledge~\cite{jin2019what}. 
Baltes \etal{}~\cite{DBLP:conf/msr/BaltesDT008} observed that comments have a closer temporal relationship with edits than posts (i.e., a question or an answer), that is, the time difference between comments and post edits are smaller compared to comments and post creations.
Typically, this happens for clarification purposes as answers and discussions shed new light on the original problem.  
For example, sometimes the asker's question may not include enough technical detail to be easily answered; similarly, the asker may conflate several issues into one posting.  
In these cases, the asker may seek to clarify the content of their question by adding new context or editing out extraneous details.  
Also, sometimes new answers emerge to older questions as the accompanying technologies evolve.
Thus, it is important to recognize that the question discussions can affect the evolution of the question itself; the question version that appears to a casual reader may have evolved since its original posting.

In this RQ, we study how question discussions are associated with the evolution of questions. 
More specifically, we study the association between the number of comments and question revisions; we do so to better understand how question discussions affect the evolution of the question content. 
We also study the association between the number of comments and the \emph{answer-receiving-time} to explore how question discussions affect the Q\&A process.
%


\rheader{Approach:}%
To understand how question discussions affect the evolution of questions, we first study the correlation between question discussions and question revisions. 
Here, we are mainly interested in the scale of question edits in terms of the size of question content change in the question body.  
Specifically, we calculate the change in the number of characters in the question body between its initial version and the current version.
We also categorize all questions into three groups, i.e., questions with no discussions (\Qnd{}), questions with comments (\Qd{}), and questions with chat rooms
(\Qwc{}). 
For each question from any category, we calculate the character length difference between the current version of the question and its initial version to investigate how question discussions are associated with the changes in the question content over a question's lifetime.

To understand how question discussions associate with the speed of question answering, we study the correlation between the number of received comments before answering activities and the \emph{answer-receiving-time}. Similar to RQ1, here we investigate the \emph{answer-receiving-time} of two different answering events: the \emph{answer-receiving-time} for the first answer (i.e., \tfa{}) and the \emph{answer-receiving-time} for the accepted answer (i.e., \taa{}).
For each question, we compute both \tfa{} and \taa{}.
We then group the questions by the number of received comments before the first answer and accepted answer respectively. 
Finally, we measure the Spearman correlation\cite{spearman1961proof} between the number of comments and the median \tfa{} (\taa{}) for questions with the same number of received comments before the first answer (accepted answer) is posted.

\begin{figure}[ht]
    \centering
    \includegraphics[width=\textwidth]{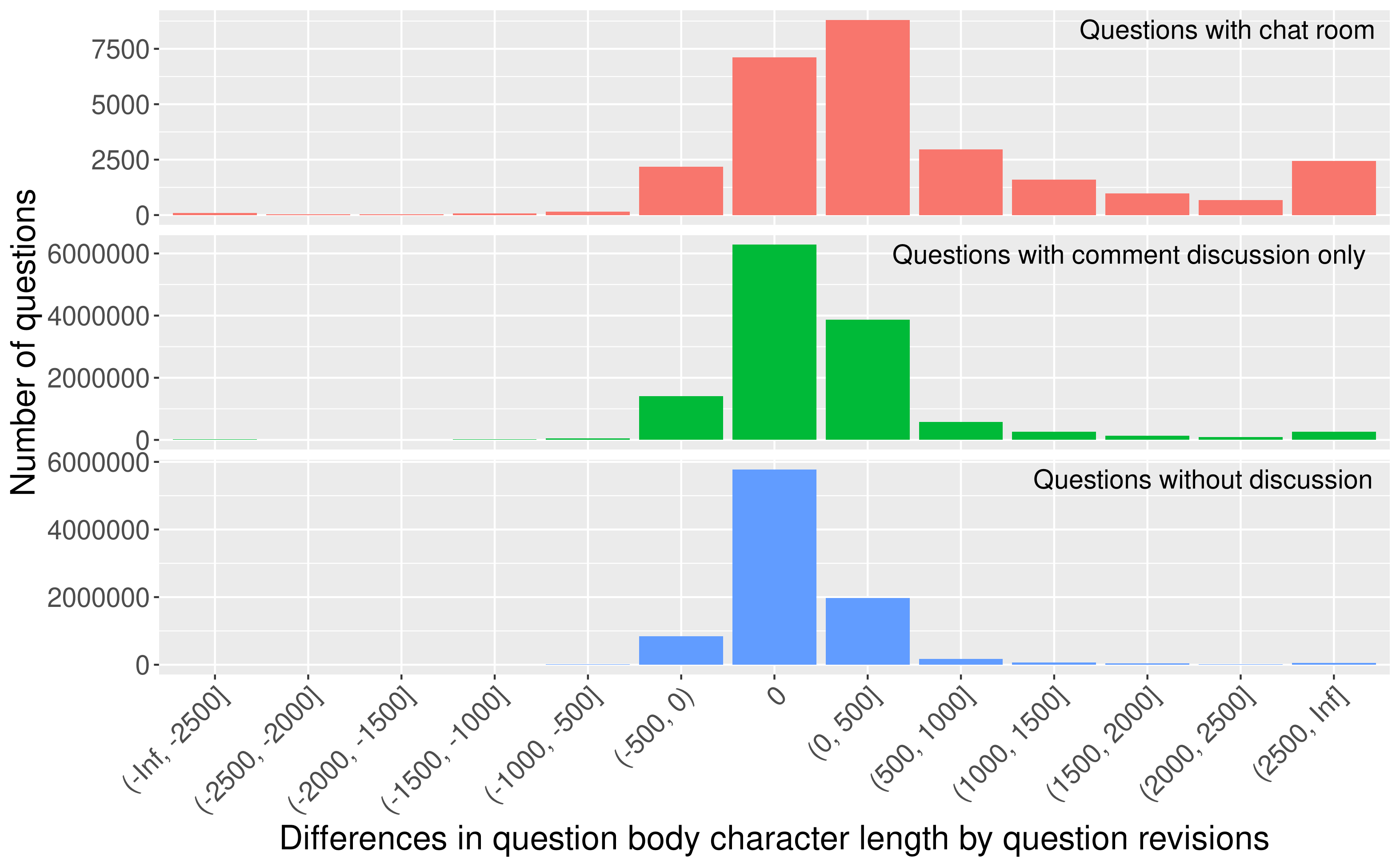}
    \caption{The distribution of the number of questions to the change in question body character length after the question is posted at different levels of question discussion activity}\label{fig:question_edits}
\end{figure}

\rheader{Results:} 
\textbf{Questions with chat rooms are more likely to be revised than questions without chat rooms, with a median size increase of 114 characters.}
Questions without chat rooms, on the other hand, do not exhibit a net change in size, although such questions may still receive edits.
Thus, the existence of a chat room attached to a question makes it more likely that the question will undergo significant revision.
Fig.~\ref{fig:question_edits} shows the distribution of questions by the change in question body length after the question is posted, according to different levels of question discussion activities. 
From the figure, we can observe that while \Qnd{} and \Qwc{} 
share the same median and modal of zero characters change in question body length, a higher proportion of questions with comments receive revisions that lead to an increase in the question body length.


\begin{figure}[ht]
    \centering
    \includegraphics[width=\textwidth]{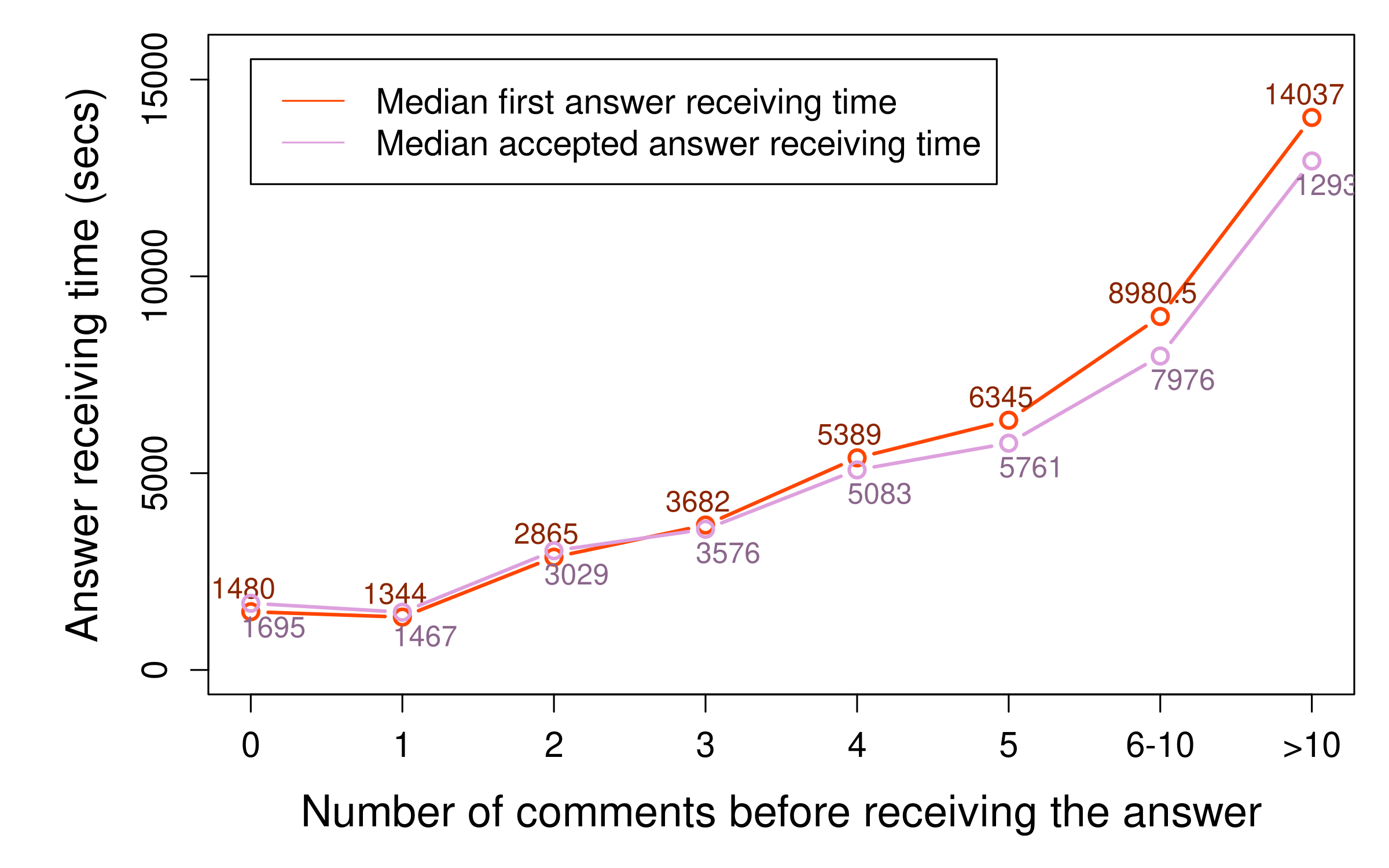}
    \caption{Median \emph{answer-receiving-time} with respect to the number of comments that are posted before the answer. The median is only calculated for questions with answers and questions with accepted answers respectively.}\label{fig:time_question_post_to_answer}
\end{figure}

Overall, \textbf{the number of comments is strongly correlated with both \tfa{} (i.e., $\rho = 0.709$, $p \ll 0.05$) and \taa{} (i.e., $\rho = 0.806$, $p \ll 0.05$)}.
Fig.~\ref{fig:time_question_post_to_answer} shows the median \tfa{} and \taa{} of questions with respect to the number of received comments before their respected answering events.
Questions with many discussions also take a longer time to answer.
One possibility is that the difficulty of these questions is also higher, therefore requiring more effort by the users to have an extended discussion before the question can be answered.
At the same time, for the \emph{answer-receiving-time} of \Qwc{}, we find that it takes a median of 5,935 secs (i.e., 1.6 hrs) and 8,438.5 secs (i.e., 2.3 hrs) to receive the first answer and the accepted answer.
The answering time follows the same trend of more discussions, i.e., a longer answering time.
The strong correlation between the number of comments that a question receives and the \emph{answer-receiving-time} suggests a close relationship between question discussions and creating answers. 
Our findings suggest that after a question is asked, interested users may offer help first in comments when an answer can't be created immediately. 
Therefore, they begin the Q\&A process by discussing with the asker through commenting. 
This is also supported by our observations in RQ1 and RQ2 where discussions mainly begin before answering and a high proportion of answerers participate in question discussions.

\begin{rqsumbox}[title={RQ3 Summary:}]
Question revisions for \Qwc{} are more likely to lead to larger edits in the question body, with a median increase of 114 characters to the question body.
A strong correlation exists between the \emph{answer-receiving-time} and number of comments before the answer, suggesting its close relationship with answering activities.
\end{rqsumbox}



\section{Implications and Discussions}\label{sec:implications_and_discussions}

\subsection{Feedback from the community}

We shared our empirical observations on Meta Stack Overflow~\footnote{\url{https://meta.stackoverflow.com/questions/416059/}}, where users share ideas and discuss the operation of Stack Overflow.
We are glad that the users from the community find our observations align with their personal experiences with question discussions on Stack Overflow.

Some users also shared their personal experiences of leveraging question discussions.
For example, one user stated ``Many questions are very localized (i.e. help only the OP (Original Poster)) and very easy to answer (i.e. obvious to anyone who has any experience). For these, writing a proper answer, with explanations, seems like a waste of time.''
It supports our theory that question discussions provide a means for alternative response than an answer.
For questions with no answers, users may still find enough information in the question discussion that can be helpful.

Other users also noticed that question discussions may be a sign of new users not realizing the edit feature for questions, ``One thing I've noticed is that new users don't seem to realize they can edit their questions. When asked for clarity, they often (attempt to) dump great chunks of information in a comment.''
The observation is supported by another user who stated ``I always add a suggestion to [edit] the question unless I am sure the user knows how to do it. Such a suggestion is not offensive, and provides the user a convenient button to edit.''
These observations also aligns with our findings that discussed questions are often edited more in RQ3.



Some users observed that comments can be deleted on Stack Overflow; future studies may wish to investigate this practice. 
Since Stack Overflow data dumps capture only a static copy of the website, researchers could monitor newly posted questions in real-time to capture deleted comments.

Another observation the community raised is that ``easy questions are often answered in comments''.
Users indicate that they find writing a quick comment can often help the asker quickly. 
However, this also introduces noise to the platform, and the reader may be uncertain where to look for such information.

\subsection{Suggestions for Researchers}

While Stack Overflow is the dominating platform for Q\&A activities for developers, Q\&A also exists in other platforms and often in other forms.
Future research can focus on the differences between Q\&A platforms to better understand the developer's preferences when asking questions.
A better understanding of developer's Q\&A needs can help us build better platforms and tools to preserve the information from the Q\&A sessions across platforms and improve the knowledge retrieval of the information for future users.

\textbf{Include discussions when modeling Stack Overflow.} Many current studies have considered asking and answering questions as isolated events.
After a question is posted, other users will read the question and try to answer it.
However, our study suggests a different story for many questions.
Discussions in the form of comments occur at large scale for questions on Stack Overflow.
The prevalence of question discussions with askers and answerers participating significantly in them suggests that they play a key role in the overall Q\&A process; consequently, any empirical study of the Stack Overflow Q\&A process has much to gain by explicitly considering question discussions in their modeling.
For example, many tools have been proposed by researchers to support developers by leveraging Stack Overflow as a knowledge base~\cite{cai2019answerbot,uddin2020mining,zagalsky2012example}.
While, these tools mined the content of questions and answers to retrieve relevant information for developers, they do not leverage the information that is contained in question discussions.
By considering question discussion in their modeling, we believe the effectiveness of these tools can be further improved with more information.

\textbf{Design automated tools to highlight information in question discussions.} Stack Overflow's overwhelming success with the international software development community is due largely to the high quality of its content, in the form of questions and answers with accompanying discussions.
However, maintaining the quality and relevance of such a large knowledge base is a challenging task; a recent study found that low quality posts hurt the reputation of Stack Overflow~\cite{6976134}. 
Because programming technologies evolve quickly, the detailed information in the questions and answers can become obsolete~\cite{8669958} and requires continual updating. 
Therefore maintaining a high quality and up to date knowledge base is very important for its users.
For this reason, Stack Overflow allows users to edit questions and answers even after a clear consensus has arisen.
Stack Overflow, as a major source of information for developers, currently does not have any mechanisms that are dedicated to the maintenance of knowledge on the platform.
Since knowledge maintenance is essential to the community, our study shows that users leverage question discussion to aid the maintenance of knowledge in the question content.
Previous studies have also observed similar phenomena in answers~\cite{8669958,soni2019analyzing}.
We suggest future research to focus on the evolution of knowledge on Stack Overflow via commenting behavior to extract best practices of the process.
By understanding the evolution of knowledge content on Stack Overflow, we can design better mechanisms on the platform to better support the community effort in maintaining knowledge.
For example, there could be report buttons for questions and answers that can raise flags regarding false information, legacy information, or potential security flaws.
Questions with such flags can be then examined by other users and therefore maintaining a knowledge base that is up to date.

\subsection{Suggestions for Q\&A platform designers}

Stack Overflow uses a gamification system based on reputation and badges to reward users who participate in the Q\&A process; for example, upvotes on questions and answers reward the original poster with reputation points.
However, at present upvotes for comments do not boost the reputation of the commenter, so their system does not currently reward participation in discussions.\footnote{\url{https://meta.stackexchange.com/questions/17364/}}
Since so much effort is put into discussions --- as evidenced by the presence of 43.6 million comments and 1.5 million chat messages in the 2021 data dump --- this seems like a missed opportunity.  
Stack Overflow could reward those users who, through their participation in discussions, help to clarify, explore, and otherwise improve the questions and answers themselves; our studies here have shown just how influential question discussions can be on improving the quality of the questions and answers.
Rewarding participation in discussions would create a positive feedback loop in the Stack Overflow gamification system, which would in turn encourage more users to engage in discussions.

\textbf{Acknowledge discussions as essential in the Q\&A, and design systems that incorporate the users' need for discussions.} A good piece of shareable knowledge starts with a good question, and Stack Overflow has practices to help ensure high quality questions.
For example, when novice users (i.e., users with newly registered accounts) first ask questions, they are led through an interactive guide on how to ask a good question. 
The guide includes both conventions (e.g., \emph{tag the question}) and best practices for asking questions (e.g., \emph{include what has been attempted to solve the question}). 
Although Stack Overflow already has a detailed walkthrough on how to ask a good question, we observed that in practice, discussing and revising questions remains commonplace.
At the same time, crowdsourced Q\&A is a labor intensive process; for example, a question may take time to attract the ``right'' answerers or a question may be hard to understand without clarification. 
In exploring RQ3, we observed that questions with extended discussions --- especially those that continue into a chat room --- tend to receive more edits to the question body.
We conjecture that question discussions can serve as a feedback loop for the asker, resulting in improvements to the questions through subsequent edits. 
Our observation also echoes a previous study which shows that tutoring novice users before they post their questions can improve the quality of their question~\cite{ford2018we}. 
We wonder if a question quality assurance ``bot'' might be able to leverage the question discussion data and mining the discussion patterns to further support askers in efficiently getting answers through crowdsourced Q\&A.

\textbf{Offer real-time Q\&A for urgent question asking, and encourage users to organize the information for future reading.}
Question discussions offer a means for askers and answerers to communicate with each other during the Q\&A process. 
Currently, chat rooms are triggered automatically once three back-and-forth comments occur between two users. 
However, there are cases where two users may wish to start a live conversation immediately. 
For example, traditionally in the open source community, it is suggested to ask urgent questions in an IRC channel to receive an immediate response~\cite{raymond2021how}. 
However, when users do so, the information during the Q\&A session will be buried in the IRC chat log. 
On the other hand, if a user were to ask the question on Stack Overflow, in exchange for not having an instant response, the Q\&A information will remain easily accessible by the public. 
While Stack Overflow already offers chat rooms as a means for instant and real-time communication, currently the chat room triggering mechanism in posting comments is an inefficient communication channel for such need. 
There exists a potential for users to choose between a synchronous or asynchronous discussion through chat rooms or comments, respectively. 
For example, Stack Overflow could build in a feature that allows users to indicate if they are available online, and are waiting for an answer. 
When other users see the indicator, they could directly start discussions in chat rooms, and later update the content of the question based on the discussion.
An intelligent communication channel selection bot could be designed to help users seek an effective type of communication by mining the historical data of communication preferences. 
Furthermore, a content summarization tool could be designed to extract pertinent information from both comments and chat rooms, for future users to better understand the context of the evolution of a question.


\section{Threats to Validity}\label{sec:threats_to_validity}

\noindent \textbf{External validity:} 
Threats to external validity relate to the generalizability of our findings.  
In our study, we focus on question discussions on technical Q\&A on Stack Overflow, which is the largest and most popular Q\&A platform for programming related questions. 
As a result our results may not generalize to other Q\&A platforms (e.g., CodeProject\footnote{\url{https://www.codeproject.com/}} and Coderanch\footnote{\url{https://coderanch.com/}}). 
To mitigate this threat, future work can consider studying more Q\&A platforms.

Another threat is that the studied Stack Overflow data dump only the current copy of Stack Overflow's website data. 
For example, users are allowed to delete their comments, answers, and questions.
This means that when users delete their comments, they are expunged from the dataset, and we are unaware of how those comments might have affected the rest of the discussion.
This concern is also shared by community members as one user stated ``the majority of the comments ever posted on Stack Overflow are probably deleted.''
Meanwhile, since there is always a valid reason for a comment to be removed, another users suggested that ``it's actually good that deleted comments are not public and Stack Overflow data dumps only capture the snapshot at the time it was taken. We don't want this kind of comments (i.e., rude/abusive comments \footnote{\url{https://meta.stackoverflow.com/questions/326494/}}) to linger for more than a quarter\dots''
Since Stack Overflow releases their data dump quarterly, we perform a comparison between the data dump from Dec. 2019 and the data dump from Dec. 2021.
From the 32.9 million question comments in 2019, only 2.1\% (i.e., 689,476) comments have been deleted in the newer data dump.
So in other words, we are unable to monitor comments that were posted and deleted within the releases of two data dumps. But if the comment survived initially, it'll likely last.

\noindent \textbf{Internal validity:} 
Threats to interval validity relate to experimental errors and bias. 
Our analysis is based on the data dump of Stack Overflow from December 2021 (the comment dataset) and web crawling in December 2021 (the chat room dataset). 
While the difference between the data dump and chat room crawling is only a month, Stack Overflow as a dynamic platform is subject to change and the data itself can evolve. 
Future work can assess our observations on new data and evaluate whether our findings continue to hold over time.

\noindent \textbf{Construct validity:} 
Since the Stack Overflow data dump not include chat room-related data, we mined that data directly from the Stack Overflow website.  
This means that our crawler and the collected data may be subject to errors (e.g., crawler timeout). 
We mitigate this issue by manually checking a subset of the collected data and verified the correctness of the scripts.


\section{Related Work}\label{sec:related_work}

\subsection{Discussion activities on Stack Overflow}

While Stack Overflow is mainly a Q\&A platform, in addition to question and answering, it also has many other mechanisms to help with the Q\&A process (e.g., the gamification system through reputation points and commenting).
In our work, we consider users posting comments associated with questions as question discussions.
However, in many other works, a discussion on Stack Overflow can have different meanings.
For example, some studies~\cite{linares2014api,shcherban2020automatic} have considered the question as a discussion (e.g., the question, all its associated answers, and any comment associated with the question or its answers).
In our work, we use discussions to describe commenting activities associated with a specific post (i.e., a question or an answer).

Most previous works on Stack Overflow discussions have a primary focus on answer discussions.
Their aim is to better understand the community efforts in improving the crowdsourced knowledge on Stack Overflow.
Zhang \etal{}~\cite{8669958} conducted an empirical study to understand answer obsolescence on Stack Overflow.
In their study, comments are used as an indicator of obsolescence for their associated answer.
A follow up study by Zhang \etal{}~\cite{zhang2019reading} examined answer comments and highlighted that the information contained in the comments should not be overlooked when reading their associated answers.
After acknowledging the importance of answers, Zhang \etal{}~\cite{zhang2021comments} focused on the current commenting mechanism on Stack Overflow and observed that the current presentation of comment information is not optimal for readers.
The comment hiding mechanism on Stack Overflow only displays the top five comments with the most upvotes. However, due to most comments never receiving any upvotes, later comments, which are likely to be more informative, are hidden from readers by default.

Comments are also viewed as triggers for post updates.
Baltes \etal{}~\cite{DBLP:conf/msr/BaltesDT008} observed that post edits often occur shortly after comment posts and suggests that comments and post edits are closely related.
Based on this observation, a study by Soni \etal{}~\cite{soni2019analyzing} further analyzed how comments affect answer updates on Stack Overflow.
Their observation echoes the finding by Zhang~\etal{}~\cite{8669958} that unfortunately users do not update their answers even with comments directly suggesting so.

Compared to the current study on discussion on Stack Overflow that mostly focuses on answers from the perspective of knowledge maintenance, our study focuses on the question discussions that mainly begin and occur during the Q\&A process. In other words, previous works have focused on preserving the knowledge while our work tends to focus more on the creation of the knowledge.

\subsection{Leveraging Discussions in Software Engineering}

During software development, communication between members of the team is important for the long-term success of the project.
Online discussions are a core part of the process, especially in open source projects where developers may be scattered around the world and rely on a variety of channels to communicate with each other~\cite{7498605}.
Since the advent of ubiquitous e-mail in the 1980s, developers have used mailing lists for discussions about the projects they are working on and interested in.
Studies show that the use of mailing lists facilitates the gathering of people with similar interests, and many open source projects still run mailing lists today~\cite{5069488} (e.g., the Gnome mailing list\footnote{\url{https://mail.gnome.org/mailman/listinfo}}).
The mailing list archive is an informative resource for researchers to understand the development of the project.
Rigby \etal{}~\cite{rigby2007what} studied the Apache developer mailing list to learn about the personality traits of developers and how the traits shift during the development of the project.
Sowe \etal{}~\cite{sowe2006identifying} studied three Debian mailing lists and constructed social networks of the mailing list to investigate how knowledge is shared between expert to novice participants.

In addition to the asynchronous email exchanges, developers also use real-time communication channels such as IRC for discussions.
IRC channels are often used by open source projects as a complement to their mailing list operations (e.g., the \texttt{\#emacs} channel on Freenode exists in addition to the project's mailing list).
Shihab \etal{} investigated \textit{GNOME GTK+}~\cite{5069488,shihab2009studying} and \textit{Evolution}~\cite{shihab2009studying} IRC channels to better understand how developers discuss in IRC.
Although e-mail and IRC are still in use today, newer and more efficient platforms have also emerged to better support the need for communication.
For example, developers report bugs and feature requests on issue trackers (e.g., Jira\footnote{\url{https://www.atlassian.com/software/jira}}), and ask questions on Stack Overflow~\cite{vasilescu2014how}.
Vasilescu \etal{}~\cite{vasilescu2014how} observed that in the \emph{R} community, developers are moving away from the \emph{r-help} mailing list to sites like Stack Overflow in the Stack Exchange network since questions are answered faster there.
Prior studies examined different communication channels aiming to better understand and improve the communication among developers.
Alkadhi \etal{}~\cite{Alkadhi:2017:RDC:3104188.3104240} applied content analysis and machine learning techniques to extract the rationale from chat messages to better understand the developers' intent and the decision making process during software development. 
Lin \etal{}~\cite{Lin:2016:WDS:2818052.2869117} studied the usage of Slack by developers and noticed that bots are in discussions to help software developers.  
%

Storey \etal{}~\cite{7498605} surveyed how developers leveraged communication channels and observed that real-time messaging tools and Q\&A platforms such as Stack Overflow are essential for developing software.  
Dittrich \etal{}~\cite{6063155} studied developers' communication across different platforms and observed that real-time messaging plays a role in the communication of developers. 
Their study shows that real-time messaging tools can support the usage of other communication channels (e.g., Skype calls) and provide a means for developers to form social and trust relationships with their colleagues.  
Chatterjee \etal{}~\cite{Chatterjee:2019:ESS:3341883.3341961} analyzed characteristics of Q\&A sessions in Slack and observed that they cover the same topics as Stack Overflow.
Wei \etal{}~\cite{WeiAutomating} applied neural networks techniques on real-time messages to automatically capture Q\&A sessions. 
Ford \etal{}~\cite{ford2018we} experimented with using real-time chat rooms for the mentoring of asking questions on Stack Overflow for novice users. 
Chowdhury \etal{}~\cite{Chowdhury:2015:MSF:2820518.2820577} leveraged information from Stack Overflow to create a content filter to effectively filter irrelevant discussions in IRC channels.
%

In our study, we focus on question discussions on Stack Overflow to better understand how they facilitate the Q\&A process.

\subsection{Understanding and Improving Stack Overflow} 

Prior research investigated how developers leverage Stack Overflow and studied different mechanisms aiming to improve the design of Stack Overflow~\cite{Xia:2013:TRS:2487085.2487140,chen2018data,ZhouBounties,8485395,ford2018we}.
Treude \etal{}~\cite{Treude:2011:PAA:1985793.1985907} categorized the types of questions on Stack Overflow, and observed that Stack Overflow can be useful for code review and learning the concepts of programming. 
Wang \etal{}~\cite{8485395} studied the edits of answers and observed that users leverage the gamification system on Stack Overflow to gain more reputation points.
Prior studies also aimed to understand the quality of the crowdsourced knowledge on Stack Overflow.
For example, Srba \etal{}~\cite{7412622} observed that an increasing amount of content with relatively lower quality is affecting the Stack Overflow community.  
Lower quality content on Stack Overflow may also affect how questions are answered. Asaduzszaman \etal{}~\cite{Asaduzzaman2013} showed that the quality of questions plays an important role in whether a question receives an answer by studying unanswered questions on Stack Overflow.  
An automated system to identify the quality of posts and filter low-quality content was proposed by Ponzanelli \etal{}~\cite{6976134}. 
To improve the quality of the crowdsourced knowledge on Stack Overflow, prior studies aimed to identify artifacts with different properties~\cite{wang2018understanding,ragkhitwetsagul2019toxic,7365804,tian2013towards,vasilescu2014how,ye2017structure,ZhouBounties}.
For example, Nasehi \etal{}~\cite{6405249} examined code examples on Stack Overflow and identified characteristics of effective code examples. 
Their study shows that explanations for code examples have the same importance as code examples.  
Yang \etal{}~\cite{Yang:2016:QUC:2901739.2901767} analyzed code snippets of popular languages (C\#, Java, JavaScript, and Python) on Stack Overflow and examined their usability by compiling or running them.
Prior studies also examined various supporting processes on Stack Overflow to better understand its operation and improve its efficiency of the crowdsourced knowledge sharing process.
Chen \etal{}~\cite{chen2018data} used a convolutional neural network (CNN) based approach to predict the need for post revisions to improve the overall quality of Stack Overflow posts.
Several studies proposed approaches to automatically predict tags on Stack Overflow~\cite{Xia:2013:TRS:2487085.2487140,6624009,beyer2015synonym}.
Wang \etal{}~\cite{Wang:2014:EET:2705615.2706107,Wang:2018:EET:3211160.3211174} proposed an automatic recommender for tags based on historical tag assignments to improve the accuracy of the labeling of tags for questions.

Instead of the extensively studied artifacts on Stack Overflow (e.g., questions, answers, tags), we investigate the question discussions by an empirical study of 43.6 million comments and 1.5 million chat room messages to understand how discussions can facilitate the Q\&A process. 


\section{Conclusions}\label{sec:conclusion}

Question discussions are an integral part of the Q\&A process on Stack Overflow, serving as an auxiliary communication channel for many developers whose technical information needs are not fully met within their nominal work environment.
Question discussions occur throughout all phases of the Q\&A process, especially before questions are answered. 
In 76.2\% of \Qda{} and 80.6\% of \Qdaa{}, the question discussions begin before the first answer and the accepted answer is posted; furthermore, 19.4\% of the question discussions begin even after the accepted answer is posted.
Question discussions allow askers and potential answerers to interact and solve the question before posting an answer.
In \Qda{}, askers participate in 60.0\% (i.e., 7.8 million) of the questions discussions and answerers participate in 34.1\% (i.e., 3.6 million) of question discussions. When the number of comments increases, a higher proportion of questions are participated by askers and answerers.
The \emph{answer-receiving-time} of a question is strongly correlated (i.e., with a Spearman correlation of $\rho = 0.709$) with the number of comments a question receives before its first answer.
We believe that our study of question discussions can be leveraged in several ways to improve the Q\&A process.  
For example, an automated triaging system could suggest an appropriate communication channel; also, bots could be designed to warn about questions that seem unclear and might require further clarification.

\section* {Acknowledgments}

We would like to thank the anonymous reviewers for their insightful comments. The findings and opinions in this paper belong solely to the authors, and are not necessarily those of Huawei. Moreover, our results do not in any way reflect the quality of Huawei software products.

\section* {Declarations}
\textbf{Conflict of Interests} The authors declare that they have no conflict of interest.


\bibliographystyle{ieeetr}      
\bibliography{references.bib}   

\end{document}